# Modular electronic microrobots with on board sensor-program steered locomotion

Vineeth K. Bandari[1,2]*, Yeji Lee[1,2], Pranathi Adluri[1,2], Daniil Karnaushenko[1,2], Dmitriy D. Karnaushenko[1,2], John S. McCaskill[1,2,3]* & Oliver G. Schmidt[1,2,3,4]*

[1]*Material Systems for Nanoelectronics, Chemnitz University of Technology, 09107 Chemnitz, Germany.*
[2]*Research Center for Materials, Architectures and Integration of Nanomembranes (MAIN), Chemnitz University of Technology, 09126 Chemnitz, Germany.*
[3]*European Centre for Living Technology (ECLT), Ca' Bottacin, Dorsoduro 3911, Venice, 30123, Italy.*
[4]*International Institute for Intelligent Nanorobots and Nanosystems, Fudan University, Shanghai, China*

*Correspondence and requests for materials should be addressed to V.K.B. (email: vineeth-kumar.bandari@etit.tu-chemnitz.de), J.S.M. (email: john.mccaskill@main.tu-chemnitz.de) or O.G.S. (email: oliver.schmidt@main.tu-chemnitz.de).

## Abstract

True microrobots, in contrast with externally controlled microparticles, must harvest or carry their own source of energy, as well as their own (preferably programmable) microcontroller of actuators for locomotion, using information acquired from their own sensors. Building on recent published work [1], we demonstrate here, for the first time, that microrobotic smartlets, hitherto buoyancy divers, can also be equipped to navigate in 2D on surfaces, with on-board control responding to both sensor information and their internal electronic program. Fabricating modular microrobots, with all dimensions of 1mm and below, has been difficult to achieve because of competing demands for the limited surface area and the challenges of integrating and interconnecting the diverse functionalities of energy harvesting, actuation, sensing, communication, docking and control. A novel high density heterogeneous integration, via soft-substrate micro flip-chip bonding of custom CMOS and LED microchiplets onto fold-up polymer surfaces, compatible with roll-up isotropic ambient light harvesting, now makes this possible. Fabricating electrolytic bubble actuators on multiple cube-faces and connecting them to a custom sensor-controlled on-board microchiplet (lablet), allows the smartlets to locomote on wet surfaces, changing direction in response to both timed programmed control as well as programmed response to locally sensed signals. Such locomoted robotic microcubes can also move to and selectively dock with other modules via patterned surfaces. This is powered by ambient light in natural aqueous media on smooth surfaces.

## Main

The extensive literature on wireless marionette micro and nanorobots [2-4] employs external magnetic [5-7], ultrasound [8-9] or optical fields [10-11] to control the movement of particles with relatively simple structure and limited or no internal information processing abilities. It is clear, however, that in the absence of specific sensory feedback from individual microrobots to the control system and individually specific actuation control of each particle in an ensemble, that independent responses to local environments cannot be achieved, and that such control does not scale well with particle number either in terms of specificity or energy demands. Fully fledged microrobots, with on-board energy transduction, sensing, actuation, communication and control, that will achieve the benefits of individually responsive agents in a microrobot collective, have been consistently regarded as technologically too demanding for sub-mm scales. Just as 1 mm$^3$ was for twenty years a grand challenge barrier for smart dust [12-14], the harder challenge of complete autonomous mobile robotic function on that scale appeared out of reach – even the surface area required for energy harvesting for robot actuation appeared too large [15]. However, recent research [1] has opened up the domain of hollow autonomous electronic microrobotics, with functional interiors and exteriors, by demonstrating how to mass produce electronic actuators and sensors along with energy harvesters in 3D in connection with prefabricated crystalline microchiplets using 3D fold-up self-assembly from planar lithography. This heterogeneous soft mass-fabrication technology significantly increases the functional density and flexibility of such systems beyond the monolithic on-chip approach to microrobotics [10, 16]. The major factors enabling this breakthrough are that additional and interior surfaces can be attained using flexible self-folding or self-rolling materials [17-19] and we have learnt how to bond near microscopic rigid chiplets [20] to these flexible substrates without inhibiting the flexible folding [1]. The use of local sensor information in the on-board programmed control of locomotion is the main novel feature that we wish to focus on in this work. Such sensory feedback also enables smart motorized self-assembly, which is otherwise limited by slow Brownian motion on scales above 1 μm.

## Fabrication, and Functional Integration of Programmable Smartlets

The modular microrobots constructed for this work are multifunctional with multiple sensors and actuators as shown in **Fig. 1**. The clean room fabrication process by means of multilayer planar lithography, microchiplet bonding to the flexible polymer layers, and autonomous rolling and folding at light-patterned locations of the upper layers, peeling off from the substrate by dissolving a sacrificial layer, is similar to that described in [1] with novel extensions as described in Methods, and demonstrates the combinatorial versatility of such fold-up modular design, here with both photosensors and independently controlled electrolytic bubble generators on three internal cube faces. The parallel 3D self-folding of arrays of smartlet cubes, allowing mass production, is shown in **SI Video S1**. The microrobots in **Fig. 1** are especially designed to move in two dimensions on a smooth planar surface, unlike their predecessors, which could only dive and resurface using electrolytic buoyancy control. The new smartlet cubes locomote on a thin film of water, 0.5 mm deep on a glass surface, by releasing bubbles from under a particular cube face. Such water films can be made naturally hygroscopic by the addition of glycerine or other additives [21], for longer term operation in moderate humidity air (e.g. > 30%), or covered with a thin hydrocarbon layer to limit evaporation. The choice of the activated actuator cube face for bubble generation determines the direction of motion, as discussed further below. The bubble generation actuators are controlled by a custom CMOS microchiplet (lablet [22], dimensions 140x140x35 μm with a 58-bit program) which is placed on the top inner surface of the cube, and this microchiplet can be programmed by light pulses. The microrobot harvests ambient light power isotropically in its custom submersible organic photocell Swiss-rolls, placed orthogonally along the free edges of the cube. The microcontroller also has two super low power differential sensory inputs, described in more detail below, which result in a binary digital signal when the sensory input exceeds a threshold voltage. For these robots, the sensors are themselves photocells placed on the inner faces of the cube. Local light signals at specific locations are used as sensory input to demonstrate the sensing capabilities of the microrobot.

## Bubble-Driven Locomotion and Directional Control

The smartlet cube microrobots are capable of autonomous on-board programmed locomotion on a two-dimensional surface as shown in **Fig. 2** with its open face oriented at the bottom towards the glass surface on which it glides. To facilitate locomotion, the surface is covered with a film (ca 500μm deep) of water, so that the smartlets, protruding above the aqueous layer, rest on the glass substrate, with their hollow interiors initially filled with water held inside despite the open bottom face by surface tension. The locomotion is two dimensional in contrast with the primarily one-dimensional buoyancy-diver locomotion demonstrated in previous smartlets [1]. The observed locomotion sequence following electronic actuation of the electrolytic bubble generation electrodes is displayed in the top panel of **Fig. 2a**. Gas bubbles (with diameters in the range of 100μm) first are generated and accumulate asymmetrically within the cube, as shown in **SI Fig. S1**, restrained by their hydrophobicity to preferentially accumulate on the actuating face. During single face actuation, the bubbles are long-lived on the timescale of the motion (e.g. 13s in **Fig. 2**), and as justified quantitatively in **SI Note S1**, the high internal pressure of such ca. 100-150 μm diameter bubbles (determined by surface tension to be in the 30-20 mbar range respectively) allows bubble-bubble repulsion near the surface to both push the bubbles downward as the vertical face fills with bubbles (SI Fig. S1), and then to lift the side of the cube until bubbles are pushed off the cube onto the gliding surface. When excess bubbles are released, the pressure drops, and the smartlet tilts back to horizontal, so that during locomotion the trailing face is periodically tilted slightly higher than the leading face. This periodic lifting of the cube face is measured from the optical images in the video sequences and traced in SI Video S2. For 150 μm (one bubble diameter) displacement per tilting period (at *ca.* 5 cycles per s) a locomotion speed of 0.75 mm/s (measured 0.8 mm/s) results. Bubble release and hence locomotion may also be facilitated by the active shape changing of the self-folded micro-origami cube, under the pneumatic pressure generated from bubble generating electrode actuators.

The different bubble generating electrode (BGE) actuators are controlled by the program running on the CMOS chiplet (interior top face) directing the microrobot as shown in Fig. 2b. Three different phases of locomotion Phase-1,2,3 are determined successively by individual actuation of one of the three BGE actuators 0,1,2 on three of the four vertical faces of the cube respectively: with actuators 0 and 2 on opposite faces. The periodic actuation pattern creates bursts of microbubbles from the actuator which drive the microrobot away from the current bubble producing face. A complete face of 150μm bubbles has a buoyancy force in water of 0.4 μN which is insufficient to overcome gravity to lift the cube (0.65 μN even for a half water-filled cube and several μN for a completely water-filled cube). What pushes the bubbles downward, lifting the cube on one side by about a bubble diameter, is the formation of new bubbles. The pressures involved here (*ca.* 30 mbar) can be calculated from the air-water surface tension T (73 mN/m @ 20°C) as $2T/R$, where R is the radius of a bubble. 30 mbar acting over 150x1000 μm corresponds to a force of 60 μN, which is more than sufficient to lift the side of the cube against gravity. Surface tension may also pull the cube down with forces up to 100 μN. So, the lowest row of bubbles then escape under the gap created by the tilted cube and the next row of bubbles are generated. With a Reynolds number of ca. 0.8, the motion is still quite strongly damped with an exponential velocity decay time (for a 1 mm diameter water-filled sphere) of 0.1s. However, a clockwise smartlet rotation preceding the orthogonal velocity locomotion behavior in successive phases, recorded in Fig. 2c is not exactly as expected. Rotation can be induced by repulsive bubble-bubble forces generated by the change to actuation bubbles on a neighboring vertical face (e.g. BGE2 in Fig. 2d) causing a non-inertial torque on the cube as described in SI Note S1, possible because of the anchoring of bubbles to the surface (by local dewetting). This rotation can clearly be seen in SI Videos S3- SI Videos S4. The exponential rotational velocity damping time constants for the cube are expected to be on millisecond timescale. As seen in SI Fig. S1, changes in orientation can also initiate sudden collective bubble movements and coalescence events, which have an inertial effect on the orientation of the cube, so that changes in actuation faces can inaugurate rotations of the cube.

## Feedback Controlled Navigation

This expanded mobility enables the microrobot's onboard program to dynamically adjust motion based on sensory input, thus allowing the microrobot to follow environment-dependent trajectories. The use of sensory feedback to control locomotion, explicitly by the sensor-mediated choice of different actuated cube faces, according to its program, is a major new feature, making this device a proper microrobot. We demonstrate this in Fig. 3, with an example using the on-board photosensor of the microrobot to detect its entry to brightly lit regions of its domain. The current-voltage characteristics of the integrated photodetector, in the dark and with 1 sun natural illumination, are shown in Fig. 3a, and the angular dependence on the presence of a light feature in the environment (laser light source) in the presence and absence of the 1 sun illumination is shown in Fig. 3b. While the overall solar powering is nearly isotropic, as in previous work [1], because of the 8 orthogonal rolled solar cells, the face-integrated photodetectors show spatially selective detection even in the presence of additional sunlight. The temporal response of the sensors, tested in Fig. 3c, ca. 10x faster for rise than fall, is on the 200 μs-2 ms timescale. The microrobot navigation was preprogrammed (using light) as detailed in Fig. 3d with a 58-bit program: See [22] for a complete explanation of the programmable capabilities of the custom microchiplets employed. The program includes a specific change of actuator output patterns when the robot senses a brightly lit region, with a different first and second time responses programmed. The actuator patterns in the three locomotion phases, separated by the transitions induced when the robot enters brightly lit regions, are shown in Fig. 3d. The image sequence in Fig. 3e contains exemplary frames from SI Video S4 of the sensory switched robotic locomotion. In contrast with previous wireless or light field-directed marionette robots, controlled by external fields, the microrobots here make their own programmed decisions to navigate based on the sensory signals they pick up from the environment as well as their own program status. Thus, in the example sequence shown in Fig. 3e, the microrobot first moves to the left along the x axis using one of its actuators until it enters the first brightly lit zone. There, its on board photosensor signal exceeds the threshold level, resulting in a programmed response into a second state, which actuates instead a second bubble generating actuator (placed at right angles to the first on an adjacent lateral face of the cube). This causes the robot to change its direction of motion in the x-y plane by 90° and move off along the y axis (upwards in the figure), and out of the current light zone, hence deactivating its sensor to again be below its threshold level. Further along the y axis, the microrobot reaches a second brightly lit zone, where again its sensor is activated, causing a state change to a third locomotion state, which actuates a third orthogonally placed actuator. This causes the microrobot to again take a 90° turn and follow a locomotion path back along the x axis in the reverse direction to at the onset. SI Videos S3- SI Videos S4 show different successive actuation sequences 0-1-0 and 0-1-2 in phases 1,2,3, with the first sequence resulting in a -x, y,-x directional motion and the second in the opposite motion in phase 3 i.e. -x, y, x.

The first light zone was illuminated statically before the experiment commenced, and even before the microrobot was programmed, so that it imparts zero bits of temporal information to the robot. The light zone is a static feature of the environment, which the robot may or may not discover. Only the motion of the microrobot into this light zone results in the microrobot detecting a changed environment (like an obstacle) and it then autonomously makes a programmed decision to change its direction of motion. For the second light zone, we do only light it up after the microrobot has left the first zone, but this is simply to reduce the amount of stray light in the experiment at one time by not having two zones lit up at the same time. Like the first zone, it is effectively a constant environmental feature for the robot, lit up while the robot is far away, which the robot must navigate into to perceive. With this on-board sensing mediated environmentally sensitive navigation, as opposed to robots dependent on operator light control, such robots can act independently and in parallel in complex environments. Finally, it is important to realize that the current microrobots, with their 58-bit on board program are capable of many other programmed behaviors than the one chosen h-ere. The sensor dependent actuation is a generic programmable option in the current on-board controller, compatible with any programmed actuation pattern and not hard-wired to a particular response as in non-programmable stimuli-response active microparticles [3].

## Active Modular Assembly

One of the new features enabled by such active, sensory switched, locomotion of autonomous microrobots is the motor-assisted self-assembly of modular microrobots, demonstrated here for the first time. This is shown in Fig. 4 and illustrated further with video material in the SI. These robots were specifically patterned with hydrophobic or hydrophilic coatings (see Methods) on their exterior faces as shown in Fig. 4a, which in contrast with solid filled robot construction can be freed of all other structures by the micro-origami infolding of the smartlet cube. The resulting attractive hydrophilic-hydrophilic and repulsive hydrophobic-hydrophilic interactions between lateral faces of the smartlet are depicted schematically in Fig. 4b and c. Two complementary experimental image sequences are shown in Fig. 4d and 4e, for non-matching and matching coat patterns of smartlets respectively. The programmed locomotion trajectories, which can be sensitive to local environments, as established in Fig. 3, can bring microrobots actively into near contact, where structurally encoded (as shown in Fig. 4a) or actively electronically mediated docking decisions can be taken. The non-matching microrobots in Fig. 4d come into close contact under the active locomotion of one of the robots, but do not dock. However, in Fig.4e the matching robots bind to one another once the actively locomoted partner reaches the vicinity of the other. That the microrobots are docked to one another securely by the hydrophilic-hydrophilic interaction is evidenced by the subsequent joint locomotion shown in [Fig. 4f](#) and [SI Video S5](#). Such patterning also allows precisely controlled docking at positions offset from full face-to-face docking as evidenced by the "brick-layer" half-offset docking shown in [SI Video S6](#) achieved by patterning faces with two vertical stripes, one hydrophilic, one hydrophobic.

## Conclusion and Outlook

This work introduces a new class of modular electronic microrobots—smartlets—that bring together sensing, actuation, ambient energy harvesting, and on-board computation in a fully untethered, sub-millimeter architecture. These microrobots achieve real-time, sensor-guided feedback control of their motion using integrated light sensors and programmable CMOS chiplets, powered exclusively by ambient light captured through rolled-up organic photovoltaics. The actuation mechanism, based on electrolytic micro-bubble generation, enables robust locomotion across thin aqueous films via asymmetrically induced buoyancy. Importantly, the integration of onboard sensors allows smartlets to alter their behavior in response to environmental cues—executing pre-programmed motion trajectories and switching behaviors based on real-time input, such as the detection of local light intensity. One of the most compelling demonstrations presented here is that of motor-assisted selective self-assembly, where programmed navigation brings microrobots into docking proximity, and hydrophilic-hydrophobic surface patterning enables deterministic binding. This interaction, driven entirely by onboard decision-making and actuation, constitutes a significant step toward microrobotic collectives capable of distributed environmental sensing and adaptive morphogenesis—without the need for external magnetic, acoustic, or optical manipulation. Beyond these functional demonstrations, the work lays a clear foundation for scalability and mass manufacturability. The fabrication process is based on planar lithography and wafer-scale bonding, enabling the automated self-folding of smartlet cubes with high precision and yield. A half-sized version (500 μm edge length) of the smartlet has already been fabricated ([see SI Fig. S2](#)), and the design is readily scalable down to 250 μm using current CMOS chip technologies. The current microcontroller, built with 180 nm CMOS, can be miniaturized to sub-100 μm footprints using 65 nm or 22 nm nodes, enabling higher onboard complexity and additional functionality. While light-harvesting efficiency scales quadratically with shrinking dimensions—posing challenges due to perimeter recombination—this limitation can be mitigated by using multiple orthogonal rolls of photovoltaic material. The transparent and foldable design of the cube also allows interior and exterior surface deployment of components, maximizing space efficiency and functional integration. Modern pixelated light sensors and electrochemical detectors, already capable of interfacing with CMOS down to 100 μm scales, further support the vision of deep environmental sensing and decision-making at micro scales [24, 25]. Looking ahead, actively navigating microrobots, as opposed to relying on Brownian motion or stochastic external fields, offer dramatically improved trial rates for docking and self-assembly. With more sophisticated sensor arrays, future microrobots could autonomously search for docking partners, forage for power modules, or respond to complex stimuli in unstructured environments. This work thus opens the microscale to fully-fledged electronic microrobotics, where modularity, programmability, and environmental interactivity are achieved at scale. The use of lithographically defined foldable architectures and independently optimized chiplets enables automated bonding and integration of diverse functionalities, laying the groundwork for swarms of microrobots that are scalable, reconfigurable, intelligent, and affordable—in line with Wright's Law for cost-efficient mass production [4]. Together, these advances pave the way for future breakthroughs in adaptive microsystems, including intelligent micro-assembly, synthetic tissue engineering, collective behavior studies, and autonomous distributed sensing.

# Methods

## Fabrication of photosensors and solar cells

The fabrication of micro-organic photodetectors (µOPDs) and micro-organic solar cells (µOSCs) is performed on a polymeric platform using a unified process, where both devices are fabricated simultaneously and share multiple structural components.

*Bottom contact pattern:* The process begins with maskless photolithography using AZ 5214 E photoresist (MicroChemicals) to define the electrode geometry. The photoresist is spin-coated at 4500 rpm for 45 seconds to achieve a 1 µm thick film, followed by soft baking at 90 °C for 4 minutes. UV exposure (365 nm, 15 mW/cm²) is then applied for 5 seconds. After exposure, the samples are post-baked at 120 °C for 2 minutes, subjected to a full UV flood exposure for 30 seconds, and developed in AZ 726 MIF developer (MicroChemicals) for 45 seconds. Thin films of chromium (Cr, 10 nm) and gold (Au, 50 nm) are then sequentially deposited at a controlled rate of 0.5 Å/s using an electron-beam evaporator (Creative Vakuumbeschichtung GmbH). A lift-off step removes the unexposed photoresist and excess metal, resulting in a well-defined bottom contact layer critical for device integration and operation.

*ITO cathode:* A negative lithography process is applied again using AZ 5214 E photoresist over the electrode layer. A 100 nm thick indium tin oxide (ITO, EvoChem) film is deposited by magnetron sputtering at a rate of 0.2 Å/s (Moorfield). Lift-off in a mixture of acetone and isopropyl alcohol (Sigma Aldrich) is used to remove the photoresist and unwanted ITO. To stabilize the layer and enhance conductivity, the samples are annealed at 200 °C for 5 hours, optimizing performance for both µOSC and µOPD.

*Electron transport layer:* Zinc oxide (ZnO) is employed as the electron transport layer (ETL) and deposited via thermal atomic layer deposition (S100, Savanna), using dimethyl zinc (DMZ) as the precursor and water ($H_2O$) as the oxidant. The deposition is conducted at 200 °C with a rate of 30 Å per cycle. Afterward, the ZnO layer is patterned using positive lithography and selectively etched in 85% phosphoric acid ($H_3PO_4$, MicroChemicals). A post-deposition annealing step at 100 °C for 20 minutes improves the structural and electrical characteristics of the ZnO layer.

*Photoactive layer, hole transport, and anode layer*: The photoactive layer comprises a 1:1 blend of poly(3-hexylthiophene) and [6,6]-phenyl $C_{61}$-butyric acid methyl ester (P3HT:$PC_{61}BM$, Sigma Aldrich) dissolved in 1,2-dichlorobenzene at a concentration of 20 mg/mL. The solution is filtered through a 0.2 µm PTFE filter and spin-coated at 800 rpm for 90 seconds, followed by baking at 140 °C for 10 minutes, resulting in a ~200 nm thick film. Immersion in methanol for 10 minutes after spin-coating improves film morphology.

The hole transport layer (HTL) is formed by spin-coating a filtered solution of poly(3,4-ethylenedioxythiophene):polystyrene sulfonate (PEDOT:PSS, CLEVIOS™ F HC Solar) at 5000 rpm for 60 seconds, achieving a ~50 nm film. This layer is annealed at 120 °C for 10 minutes to improve conductivity and uniformity. All procedures are carried out in a nitrogen-filled glovebox to maintain film quality and prevent degradation.

The top electrode is realized by depositing 120 nm of gold at 0.4 Å/s using an electron-beam evaporator (Creavac), followed by patterning with positive lithography. The Au layer is etched using a commercial gold etchant (ACL2, MicroChemicals), and further structured using $O_2$ plasma etching (TEPLA) at 400 W for 5 minutes. A final lift-off step removes residual photoresist to finalize the top electrode geometry.

## Structural characterization and analysis

The thicknesses of the individual layers were precisely measured using a surface profiler (Veeco Dektak 8). Morphological features and surface topology were characterized via optical microscopy (Olympus BX5) and scanning electron microscopy (SEM) using a GAIA3 TESCAN system, operated at an acceleration voltage of 5 kV with a focused $Ga^{2+}$ ion beam.

*Electrical and Photovoltaic Characterization*: Current-voltage (I–V) characteristics of the devices were measured under standard illumination conditions (100 mW/cm²) using a solar simulator (LSE341, LOT QuantumDesign GmbH) equipped with 150–600 W arc lamps. The light intensity was calibrated using a commercial optometer (ILT2400, International Light Technologies). A source meter (Keithley 2636A) was used to record the I–V curves. The transient photoresponse of the µOPD was monitored using a digital oscilloscope (Rohde & Schwarz GmbH, RTB2004), enabling detailed analysis of photocurrent behavior under periodic light illumination. These measurements facilitated evaluation of the response time and operational stability of the µOPD devices. For angle-dependent measurements, the smartlet was mounted on a motorized rotation stage to precisely control its orientation relative to the incident light. The solar simulator was fixed perpendicular to the initial plane of the smartlet at a distance of 10 cm. By adjusting the tilt angle of the smartlet, the influence of illumination angle on device performance was systematically investigated under reproducible and controlled conditions.

*Environmental Conditions:* All measurements were conducted under ambient atmospheric conditions to ensure that the device performance reflects realistic operational environments.

**Fabrication of self-assembling polymeric platform**
The self-assembling polymeric platform forms the foundational architecture for both tubular µOSCs (micro-organic solar cells) and 3D micro-origami cubes. The fabrication process involves sequential patterning of a multilayer polymer stack on a planar substrate, followed by the initiation of self-rolling or self-folding through the controlled release of sacrificial layers. Initially, a glass substrate is treated with oxygen plasma (TEPLA) at 400 W for 2 minutes to enhance adhesion for subsequent polymer layers. A lanthanum–acrylic acid-based organometallic photo-patternable material is then spin-coated at 3000 rpm for 90 seconds, soft-baked at 40 °C for 10 minutes and exposed to UV light using a MA6 Mask Aligner (Süss, 365 nm, 15 mW/cm²) for 60 seconds. The film is developed in deionized (DI) water for 10 seconds and hard baked at 220 °C for 30 minutes, forming a sacrificial layer (SL) with a thickness of approximately 300 nm.

Next, a photo-patternable hydrogel layer (HGL)—which functions both as the hinge for origami cube folding and as the rolling base for tubular µOSCs—is spin-coated at 6000 rpm for 90 seconds. After a soft bake at 40 °C for 10 minutes, it undergoes UV exposure (365 nm, 15 mW/cm²) for 90 seconds using the same mask aligner. The exposed film is developed in diethylene glycol monoethyl ether (Sigma Aldrich) for 20 seconds and hard baked at 220 °C for 30 minutes to yield an ~800 nm thick cross-linked hydrogel layer.

A photo-patternable polyimide is subsequently applied as the rigid structural layer. This is spin-coated at 5000 rpm for 90 seconds, soft-baked at 50 °C for 10 minutes, and exposed to UV light (365 nm, 15 mW/cm²) for 70 seconds. Development is performed using a solution consisting of 1 wt% ethanol, 2 wt% diethylene glycol monoethyl ether, and 4 wt% 1-ethyl-2-pyrrolidone (all Sigma Aldrich) for 60 seconds. A hard bake at 220 °C for 30 minutes finalizes the layer, producing a polyimide film with ~500 nm thickness. The final layer—a 10 µm thick SU-8 25 structural reinforcement layer (Micro Resist Technology)—is spin-coated at 3000 rpm for 60 seconds. It is soft-baked at 95 °C for 4 minutes, exposed to UV light for 80 seconds, post-baked at 95 °C for 2 minutes, and developed in mr-Dev 600 for 60 seconds. This square-shaped SU-8 structure provides robust support and forms the rigid base of the 3D cube during folding and further manipulation.

Following the complete formation of the polymeric platform, µOSCs are fabricated directly on the planar stack, as described in the previous section.

*Micro-origami self-assembly process*: The transformation of the planar polymeric platform into tubular or cubic microstructures is triggered through a solution-based self-assembly process. This process is initiated by selective etching of the sacrificial layer and is finely tuned to control rolling and folding dynamics. A 0.1 M aqueous solution of sodium diethylenetriamine pentaacetic acid (DTPA, Sigma Aldrich) is prepared, and the pH is carefully adjusted from 6 to 9 using sodium hydroxide (NaOH, Sigma Aldrich). This pH adjustment is critical for regulating the etch rate and selectivity during sacrificial layer removal, which directly affects the precision and reproducibility of the rolling and folding mechanisms. Upon initiating the etch, the planar structure undergoes self-assembly into either tubular or cubic configurations depending on the pre-patterned design. Once the transformation is complete, the structures are immersed in deionized (DI) water for 20 minutes to remove any residual rolling agents. Gentle agitation during this rinse step helps ensure structural integrity while eliminating remaining chemical residues. To maintain the shape and functionality of the assembled microstructures, the smartlets are stored in medical-grade saline or phosphate-buffered saline (PBS) solution until further processing or testing.

**Preparation for SLID bonding**

To enable chip integration on the smartlet, Solid-Liquid Interdiffusion (SLID) bonding is employed using copper (Cu) and tin (Sn) as the conductive and soldering materials, respectively. These metals are chosen for their well-established reliability in microelectronic bonding applications. Cu and Sn layers are electroplated using standard techniques, with final layer thicknesses of approximately 10 μm for Cu and 5 μm for Sn. After deposition, a reflow process is performed under vacuum to initiate interdiffusion and form uniform solder bumps. This step enhances bonding strength and electrical contact quality. The alignment and bonding of flipped chips are carried out using a commercial flip-chip bonder (FinePlacer Pico). Bonding parameters such as pressure, temperature, and alignment are precisely controlled to ensure accurate positioning and robust electrical and mechanical interconnection between the microchip and the underlying structure.

**Lablet μChips**

The "lablet" micro-chiplets were obtained by thinning and singulating individual dies from combinatorial variant array chips, each containing approximately 1000 distinct lablets. These chips were designed at Ruhr Universität Bochum and fabricated at wafer scale using a 180 nm CMOS process by TSMC. Each lablet features two contact pads that function as a 1-bit digital input for interfacing with external sensors. The internal digital logic is programmed to respond to this input according to one of eight predefined logical conditions. Upon detection of a valid input, the lablet initiates a specific programmed action. This action may involve transitioning from Phase 1 to Phase 2, Phase 2 to Phase 3, or entering a loop mode in which the sequence of phases is repeated continuously until interrupted. A detailed description of the internal architecture, logical conditions, and state transition design is provided in [22].

**Fabrication of BGEs**

The complete fabrication process of a smartlet begins with constructing the polymeric origami platform, followed by the sequential integration of the bubble-generating electrodes (BGEs), μOSC and μOPD deposition and patterning, SLID bonding of the μChip, and final passivation using SU-8 photoresist. The BGEs are strategically patterned on three cube faces to enable directional actuation and controlled locomotion.

*Oxygen Evolution Reaction:* To enable oxygen evolution, a thin nickel (Ni) film is deposited due to its known photocatalytic activity. Ni is patterned using negative lithography and deposited via magnetron sputtering at a controlled rate of 0.5 Å/s to achieve a final thickness of 15 nm. After deposition, a lift-off process using acetone and isopropanol defines the patterned Ni regions by removing excess material and unwanted residues.

*Hydrogen Evolution Reaction:* Platinum (Pt) is employed as the primary catalyst for the hydrogen evolution reaction. As with the Ni process, negative lithography is used to define the Pt deposition regions. A titanium (Ti) adhesion layer is first deposited using an electron-beam evaporator (Creavac) at 0.2 Å/s to a thickness of 5 nm, followed by a 10 nm Pt layer deposited under the same conditions. The resulting Ti/Pt stack is patterned using a lift-off step in acetone and isopropanol, yielding precisely defined catalytic surfaces for HER.

**Fabrication of hydrophobic layer for patterned smartlet docking control**
This procedure is applied exclusively for customizing the exterior faces of smartlet cubes to enable selective docking behaviour. The process begins with AZ 5214E negative lithography (as described above) applied to the sacrificial layer (SL) to define areas corresponding to specific cube faces. A 200 nm thick amorphous molybdenum trioxide ($MoO_3$) layer is then deposited using electron-beam evaporation (Edwards) at a rate of 0.2 Å/s. Following deposition, a lift-off process is performed in acetone and isopropanol to remove unwanted material and expose patterned $MoO_3$ regions. This $MoO_3$ layer acts as a water-soluble sacrificial template for the hydrophobic material patterning. The hydrophobic coating material (ETC-PRO, EVOCHEM) is deposited to a thickness of 100 nm using electron-beam evaporation (Edwards). After deposition, the sample is immersed in deionized (DI) water, which selectively dissolves the $MoO_3$ layer. As a result, the hydrophobic material deposited on top of $MoO_3$ is lifted off, while the material directly on the substrate remains, forming a clean and well-defined hydrophobic pattern. Finally, the samples are cured by heating at 170 °C for 2 hours to activate and stabilize the hydrophobic surface properties.

**Experimental setup for controlled locomotion and self-assembly**
The experimental setup consisted of a 70 mm × 70 mm glass substrate placed on a soft tissue layer to minimize external mechanical disturbances. Prior to use, the glass was treated with oxygen plasma to render the surface hydrophilic, facilitating uniform water film formation. A ~500 μm thick water layer was created by slowly dispensing water droplets onto the treated surface. The smartlet was carefully picked up, flipped such that its open face faced downward, and gently placed onto the water-covered glass substrate. Manual alignment was performed to orient the μOPD toward the left side of the experimental field. All experiments were recorded using a Keyence VHX digital microscope camera. To power the smartlet, a solar simulator calibrated to 1 sun intensity was positioned at a 45° angle from the right side of the substrate, ensuring homogeneous illumination. Additionally, a high-power white LED was used for optical programming. This LED was electronically connected to a logic analyzer, which controlled its on/off state according to the programmed sequence. The analyzer synchronized the light pulses with the program commands encoded as binary input bits.

In experiments without feedback control, the solar simulator was activated to power the smartlet, enabling its onboard lablet μChip. A predefined optical signal sequence was delivered from the LED, consisting of an 8-bit program command followed by a 58-bit run command. This sequence was detected by the smartlet's μOPD and interpreted by the internal digital logic, which then triggered the corresponding behavioral routine. For feedback-based experiments, a 635 nm laser was directed perpendicularly onto the substrate from above. Once the smartlet received the run command and became active, it moved toward the laser spot. Upon detecting the laser illumination through its μOPD, the smartlet reoriented itself, altering its behavior accordingly. The laser spot was then moved to a new location, prompting the smartlet to follow it. This cycle demonstrated real-time optical feedback control and adaptive behavior in response to spatially localized stimuli. The smartlet's locomotion is powered by a self-propulsion mechanism involving cyclic bubble generation and expulsion. This mechanism consists of face deformation, bubble formation, bubble release, and structural recovery— each phase contributing to net displacement across the water surface (as illustrated in Figure 2). To improve visibility and controllability of bubble formation, methylene blue (0.075%) was added to the water film. The dye reduced bubble adhesion and enhanced detachment, preventing accumulation at the actuation site.

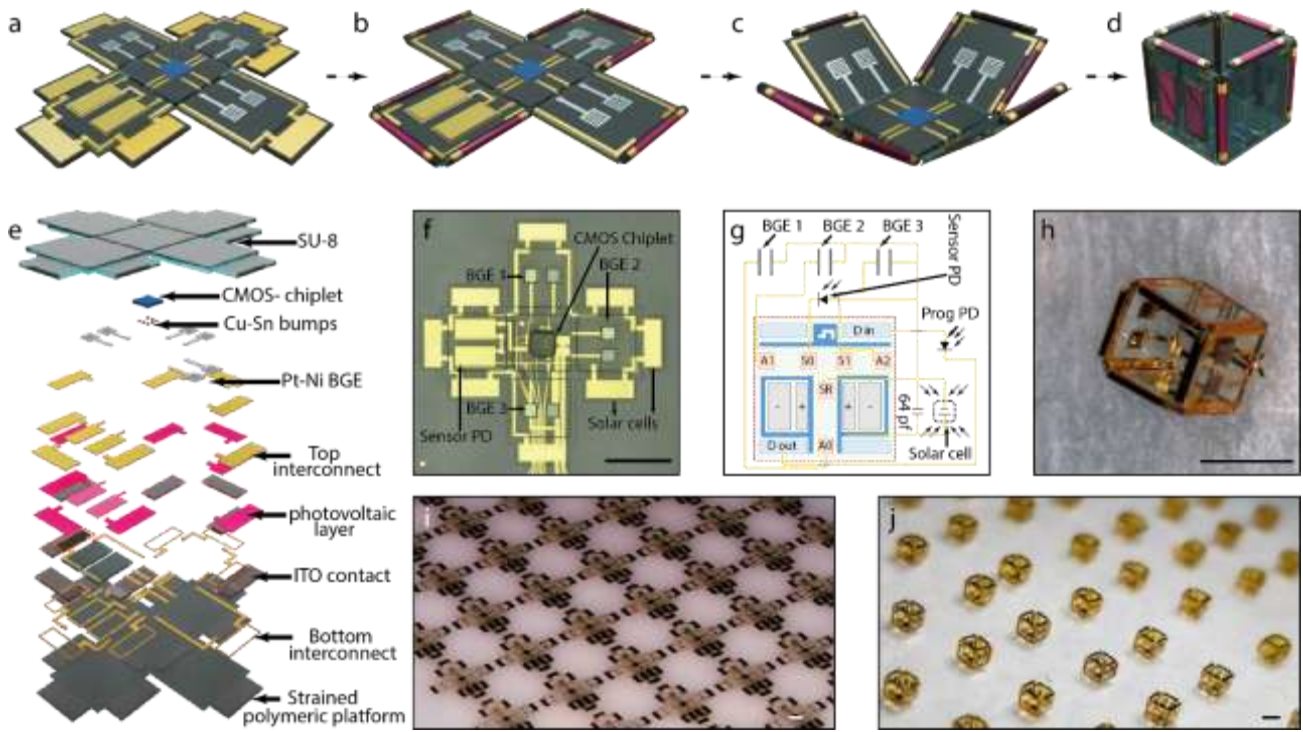

**Figure 1. Design, fabrication, functional integration, and self-assembly of fully autonomous smartlet microrobots with multiple bubble actuators. a–d,** Sequential folding process of a programmable micro-origami cube (smartlet). **a,** Initial planar structure with integrated functional layers. **b,** Rolling of peripheral flaps containing photovoltaic modules. **c,** Intermediate folding stage during origami-based self-assembly. **d,** Final folded configuration forming a 3D robotic cube. **e,** Exploded schematic of the fabrication stack. The structure comprises a pre-strained polymeric base, patterned gold bottom interconnects, an ITO contact layer, organic photovoltaic and photodetector (PD) layers, top gold interconnects, platinum–nickel bubble-generating electrodes (BGEs), Cu–Sn solder bumps for SLID bonding to a CMOS chiplet, and an SU-8 encapsulation layer for mechanical and chemical protection. **f,** Optical micrograph of a fully fabricated smartlet in its unfolded (planar) state, highlighting chiplet integration and multilayer patterning. Scale bar, 500 μm. **g,** Circuit schematic of smartlet functionality. Three independently addressable BGEs (BGE1–3) are connected to actuation pads (A0–A2) on the 140x140x35 μm CMOS chiplet (lablet). A programmable PD interfaces with sensor and logic inputs/outputs (Din, Dout), while the integrated solar cell connects to dedicated power pads and charges a 64 pF capacitor for energy storage and power conditioning. **h,** Optical image of a self-assembled smartlet cube. Scale bar, 500 μm. **i,** Wafer-scale fabrication of smartlets on a 2-inch substrate, showing highly parallel processing prior to self-assembly. Scale bar, 3 mm. **j,** Array of fully self-assembled smartlets after release and folding. For full time sequence, see SI Video S1. Scale bar, 3 mm.
.

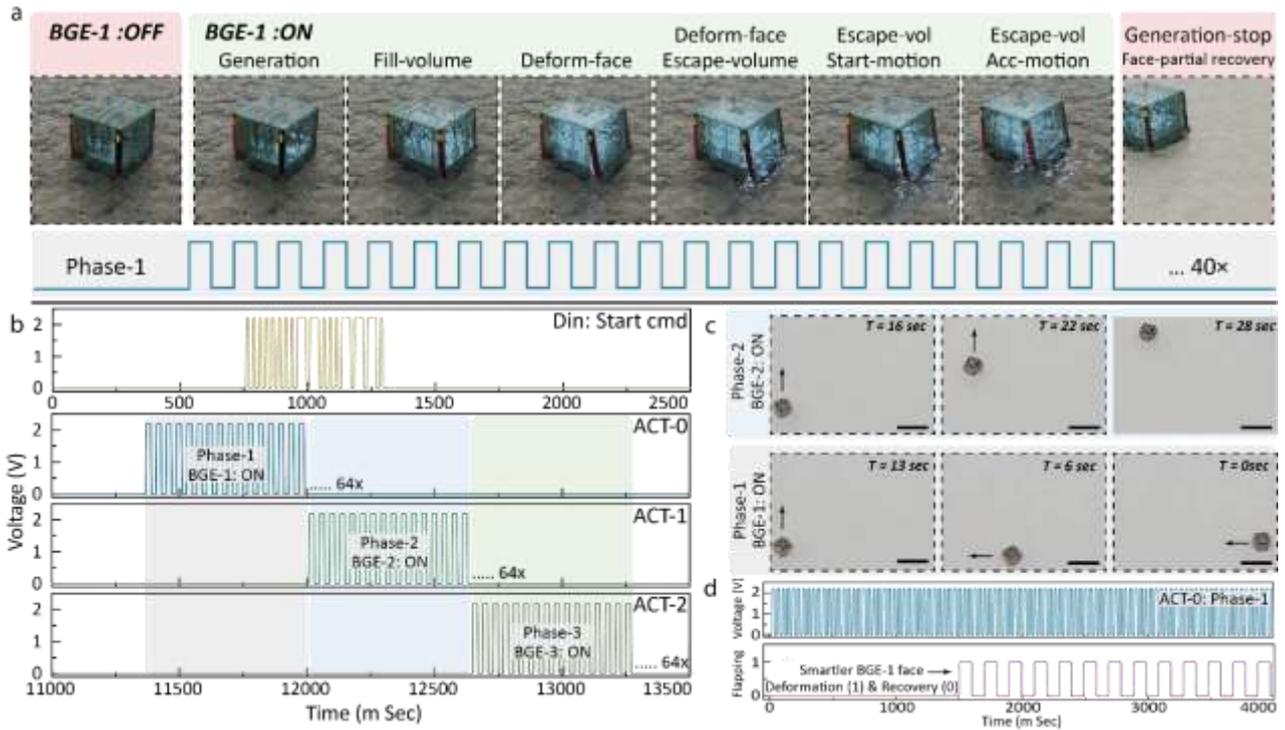

**Figure 2 CMOS-programmed actuation and controlled directional locomotion of smartlets via programmed bubble propulsion. a,** Schematic illustration of the locomotion mechanism enabled by BGE1. The smartlet undergoes a programmed actuation sequence (Phase 1, bottom) where bubble generation increases internal pressure, deforming the BGE1-containing face and creating an escape gap. As bubbles exit, thrust is generated, causing net displacement. After the actuation pulse ends, the face recovers and reseals, completing one propulsion cycle. **b,** CMOS chiplet run command and actuation signals. Actuation signals ACT-0, ACT-1, and ACT-2 correspond to the activation of BGE1, BGE2, and BGE3, respectively. Only one actuator is active during each phase: Phase 1 activates ACT-0 (BGE1), Phase 2 activates ACT-1 (BGE2), and Phase 3 activates ACT-2 (BGE3), enabling programmable directional control. **c,** Time-lapse optical images showing phase-specific locomotion. In Phase 1 (bottom), activation of BGE1 propels the smartlet leftward. In Phase 2 (top), activation of BGE2 results in upward movement. Movement direction is determined by the spatial orientation of the active BGE. For full time sequence, see SI Video S2. Scale bars, 2 mm. **d,** Top: CMOS-generated actuation pulses for Phase 1 (ACT-0). Bottom: Corresponding binary trace of smartlet face tilting (high = tilted, low = recovery), confirming dynamic actuation and relaxation behaviour in real time.

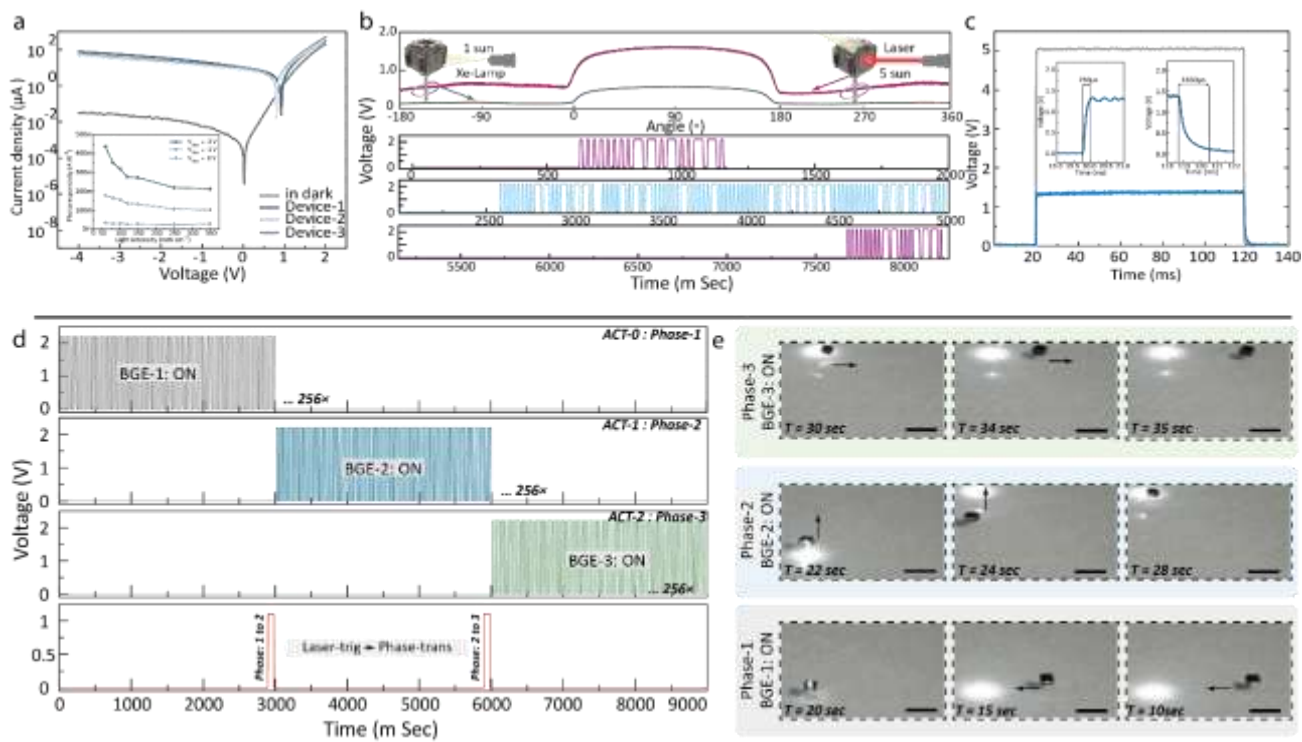

**Figure 3 | Photodetector performance and light-triggered directional locomotion of smartlets. a,** Current density–voltage (J–V) characteristics of the integrated photodetector (PD) measured in the dark and under 1 sun illumination for three representative devices. Inset: Photoresponsivity as a function of light intensity at bias voltages of 0 V, –1 V, and –2 V. **b,** Top graph: Angular dependence of PD output voltage under illumination. The lower curve represents 1 sun intensity from a xenon lamp; the upper curve combines 1 sun global illumination and 5 sun laser illumination. The light source was positioned at a distance from the device, with 90° corresponding to normal incidence on the PD. Second-Fourth graph: CMOS programming signals showing: second — program command sequence (Din), third — 58-bit Manchester encoded instruction stream, and fourth — run command initiating actuation cycles. **c,** Transient PD response under pulsed illumination. The gray trace shows the input signal; the blue trace represents the PD output. Insets: Expanded views of the rise (230 μs) and fall (1850 μs) times, confirming fast light responsiveness. **d,** CMOS-generated actuator signals for BGE-1, BGE-2, and BGE-3 (top to bottom), each corresponding to a distinct propulsion phase. The lower trace indicates phase transitions triggered by laser pulses detected by the PD. The first laser event switches from Phase 1 to Phase 2, and the second from Phase 2 to Phase 3. **e,** Time-lapse frames demonstrating directional control of smartlet locomotion. Activation of BGE-1 induces leftward drift (bottom), BGE-2 drives upward motion (middle), and BGE-3 causes rightward propulsion (top). Motion direction corresponds to the active actuator and is guided by spatially modulated light detection. For full time sequence, see SI Video S4. Scale bars, 2 mm.

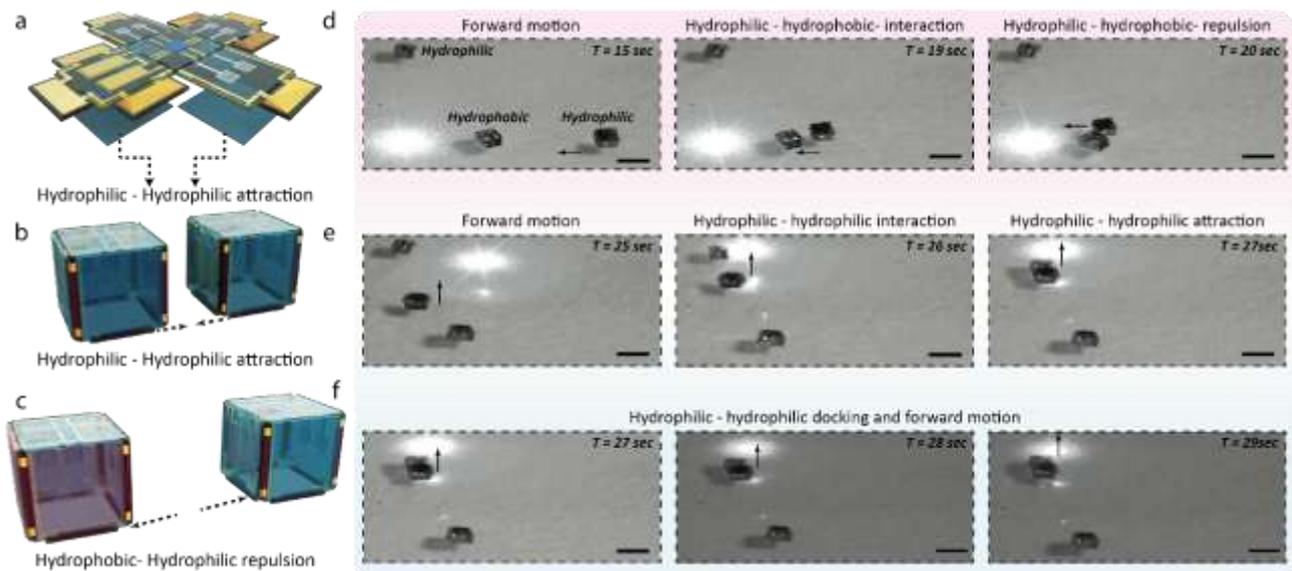

**Figure 4 | Surface chemistry–mediated docking and de-docking in modular smartlets. a,** Planar smartlet design with designated outer faces coated with either hydrophilic or hydrophobic materials. **b-c,** Repulsion of non-matching and attraction of matching hydrophilic and hydrophobic smartlet faces in water. **d,** Time-lapse frames capturing repulsion during locomotion: a hydrophilic smartlet approaches a hydrophobic one (left), initiates contact (middle), and is repelled due to mismatched hydrophilic and hydrophobic surfaces (right). Scale bar, 2 mm. **e,** Time-lapse sequence showing docking via hydrophilic–hydrophilic attraction. As hydrophilic smartlet moves upward (left) and nears another hydrophilic smartlet (middle), the pair adheres via capillary or surface tension forces (right). Scale bar, 2 mm. **f,** Two docked smartlets remain stably connected while moving together as a unit, demonstrating robust docking integrity during propulsion. For full time sequence, see SI Video S5. Scale bar, 2 mm.

# Modular electronic microrobots with on board sensor-program steered locomotion

## Supplementary information

Vineeth K. Bandari[1,2]*, Yeji Lee[1,2], Pranathi Adluri[1,2], Daniil Karnaushenko[1,2], Dmitriy D. Karnaushenko[1,2], John S. McCaskill[1,2,3]* & Oliver G. Schmidt[1,2,3]*

[1]*Material Systems for Nanoelectronics, Chemnitz University of Technology, 09107 Chemnitz, Germany.*

[2]*Research Center for Materials, Architectures and Integration of Nanomembranes (MAIN), Chemnitz University of Technology, 09126 Chemnitz, Germany.*

[3]*European Center for Living Technology (ECLT), Ca´ Bottacin, Dorsoduro 3911, Venice, 30123, Italy.*

*Correspondence and requests for materials should be addressed to V.K.B. (email: Vineeth-kumar.bandari@etit.tu-chemnitz.de), J.S.M. (email: john.mccaskill@main.tu-chemnitz.de) or to O.G.S. (email: oliver.schmidt@main.tu-chemnitz.de)

**The PDF file includes:**

Supplementary information (SI) **Figures S1-S2**

Supplementary information (SI) **Note S1**

Other Supplementary information for this manuscript include the following:

**Video S1.** Micro-origami enabled 2D to 3D transformation of the smartlet

**Video S2.** CMOS-programmed actuation and controlled directional locomotion of smartlets

**Video S3.** Feedback controlled navigation enabled by integrated photo-sensor & μ-chiplet: Two-point sensing with leftwards, straight and leftward trajectory

**Video S4.** Feedback controlled navigation enabled by integrated photo-sensor & μ-chiplet: Two-point sensing with leftwards, straight and rightward trajectory

**Video S5.** Active μ-modular assembly enabled by sensor-guided locomotion in smartlets

**Video S6.** Site selective deterministic smartlet docking enabled by patterned surface modification

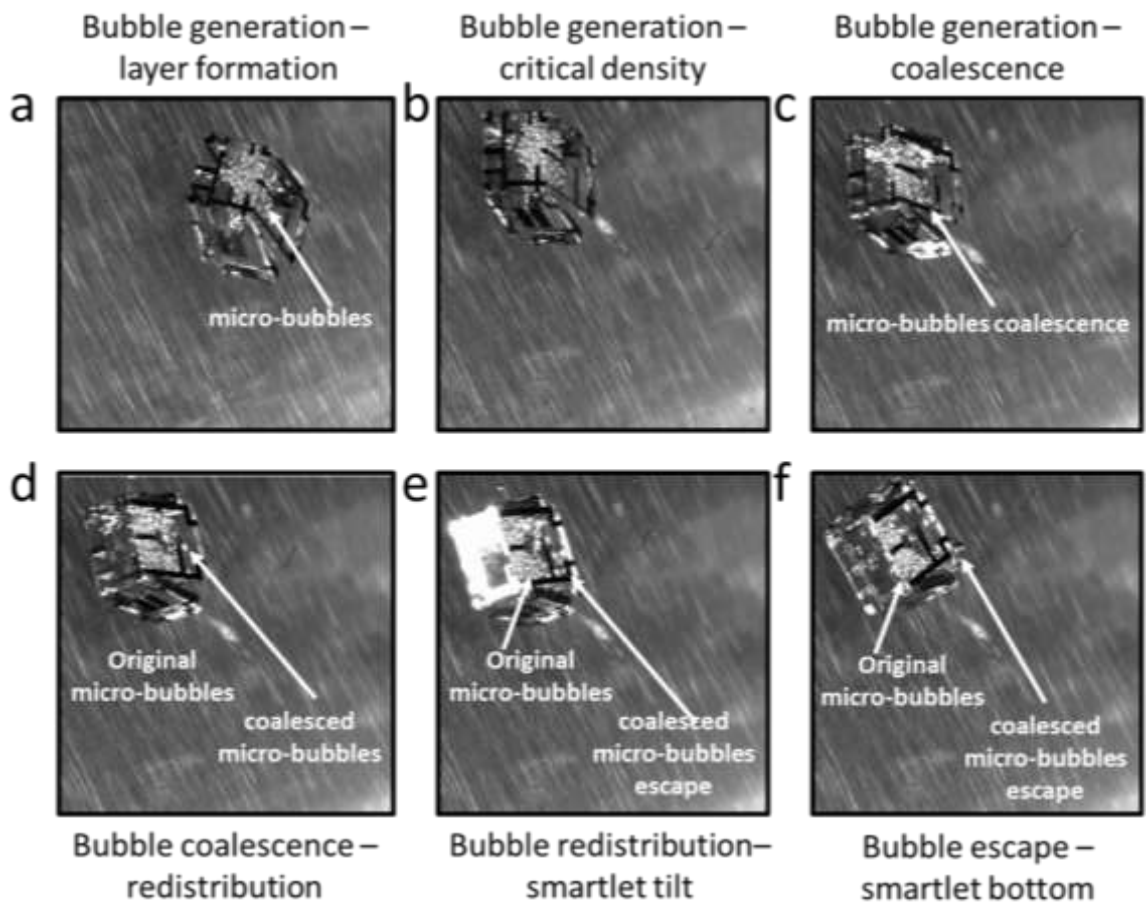

**Figure S1 | Electrolytic bubble packing and release dynamics during smartlet actuation. a)** Electrolysis-driven generation of microbubbles inside the smartlet. **b)** Accumulation of bubbles and rise in internal bubble density toward a critical threshold. **c)** Onset of bubble coalescence as the critical density is exceeded. **d)** Redistribution of coalesced bubbles within the smartlet interior. **e)** Asymmetric bubble distribution induces a shift in the center of buoyancy, triggering smartlet tilting. **f)** The tilting of smartlet causes the release of bubbles through the bottom opening.

**Supplementary Note 1: Mechanistic analysis of smartlet tilting and locomotion induced by bubble formation**

The locomotion of the smartlet microrobots emerges from a finely tuned balance between bubble formation through electrolysis, surface tension effects, and the interaction between the cube's shape and its aqueous environment. This supplementary note provides a quantitative and mechanistic analysis of the forces at play in lifting, tilting, and propelling the smartlet.

***Physical Configuration:*** Each smartlet is simplified here to a hollow 1 mm³ cube with walls ca 40 µm thick one face open, not considering the fine µOSC rolls on 8 of its edges (see Fig. 1). During experiments, it is placed with the open face downward on a thin water film approximately 0.5 mm thick, allowing water to partially flood the internal cavity. Depending on the degree of filling—completely water-filled, filled to the waterline, or gas-filled—the effective force required to initiate tilting changes significantly. The weight of the smartlet is

***Calculated Gravitational and Buoyant Forces:*** Let us consider three scenarios of internal content distribution:

1. ***Fully Water-Filled Smartlet (Half-Submerged)***
*A. Ideal thin-walled cube with no weight in walls*
  - Cube volume $V_c \approx 1 \times 10^{-9}$ m³
  - Submerged volume $V_s \approx 0.5 \times 10^{-9}$ m³
  - Gravitational force: $F_g \approx \rho (V_c - V_s) g \approx$ 1000 kg/m³ × 0.5e-9 m³ × 9.8 m/s² ≈ 4.9 µN
  using Archimedes' principle for the displaced volume of water buoyancy force.

*B. On average 40 µm thick walls with measured smartlet weight*
  - Interior volume $V_i \approx 0.81 \times 10^{-9}$ m³,
      above $V_{iu} \approx 0.39 \times 10^{-9}$ m³ and
      below $V_{il} \approx 0.42 \times 10^{-9}$ m³ waterline
  - Gravitational force on dry cube based on measured weight ≈ 0.4 mg ≈ 3.9 µN
     includes chiplets, traces and µOSCs
  - Gravitational force on water above waterline $V_{iu} \approx F_{giu} \approx \rho V_{iu} g \approx$ 3.8 µN
  - Buoyancy (displaced water by submerged walls) -1*(0.5-0.42)*9.8 = -0.8 N
  - Net gravitational force: $F_g \approx$ 3.9-0.8+3.8 ≈ 6.9 µN

2. ***Partially Water-Filled Smartlet (Filled to Waterline at 0.5 mm)***
*A. Ideal thin-walled cube with no weight in walls*
  - Gravitational force: $F_g \approx$ 0 µN
*B. On average 40 µm thick walls with measured smartlet weight*
  - Net gravitational force: $F_g \approx$ 3.9-0.8 ≈ 3.1 µN

3. ***Gas-Filled Smartlet (Half-Submerged):***
*A. Ideal thin-walled cube with no weight in walls*
  - Gravitational force: $F_g \approx \rho (0 - V_s) g \approx$ - 4.9 µN
*B. On average 40 µm thick walls with measured smartlet weight*
  - Net gravitational force: $F_g \approx$ 3.9-4.9 ≈ -1.0 µN

For tilting the smartlet along one side, only half these forces are relevant due to torque acting on the cube's edge.

***Surface Tension and Bubble Pressure:*** Bubbles formed via electrolysis grow to 50–100 µm in radius and initially adhere to the hydrophobic interior of the actuated cube face. Due to surface tension (~72.75 mN/m for water), these bubbles remain uncoalesced, forming a monolayer that exerts localized internal pressures. Using the Laplace equation:

$$P = 2T / r$$

For a typical surface tension of water T = 0.072 N/m, the internal pressures become:

- For r = 50 μm: P ≈ 29.1 mbar

- For r = 100 μm: P ≈ 14.4 mbar

These pressures vastly exceed the gravitational pressures resulting from the above calculation

$F_g/A < 7$ μN/ 1 mm$^2$) = 7 Pa = 0.07 mbar,

allowing bubble-bubble repulsion (see text) to lift one cube edge even against a gravitational loads well above 7 μN.

Note that the meniscus formed on the outside of the half-submerged cube will also contribute significant additional forces.

***Dynamic Tilting and Torque-Induced Motion:*** As bubble growth continues on the first actuated face (e.g., BGE1), accumulated internal pressure causes tilting of the cube at the edge nearest the bubble base. Upon switching actuation to a second face (BGE2), newly forming bubbles displace the residual BGE1 bubbles, creating an internal torque due to uneven bubble packing. Additionally, surface dewetting and bubble pinning at the cube's base reinforce the rotation toward previously dewetted hydrophobic patches, reducing potential energy. This dynamic leads to periodic tilting and rebalancing of the cube, causing net translational movement—a mechanism analogous to a ratchet rather than continuous sliding.

***Displacement Estimates and Drag Considerations***: Optical analysis and high-speed imaging (Fig. 2d) show a translational displacement of ~150 μm per cycle at ~5 Hz, resulting in:

$$v = \Delta x \times f \approx 0.15 \text{ mm} \times 5 = 0.75 \text{ mm/s}$$

This agrees with the measured speed (~0.8 mm/s). Drag force estimates using Stokes' Law:

$$F\_drag = 6\pi\eta Rv \approx 7.5 \text{ nN} (\eta = 1 \text{ mPa·s}, R = 0.5 \text{ mm})$$

This drag is negligible compared to surface tension forces (up to 60 μN), enabling sustained ratcheting motion.

***Conclusion***: While buoyant and gravitational forces are consistent with the static equilibrium of the smartlet, resting on the glass surface unless >50% filled with gas, surface-tension-induced pressures from confined, non-coalescing bubbles dominate dynamic tilting and locomotion. The synergy between asymmetric actuation, internal bubble rearrangement, and hydrophobic surface effects enables controlled 2D locomotion in the microrobot.

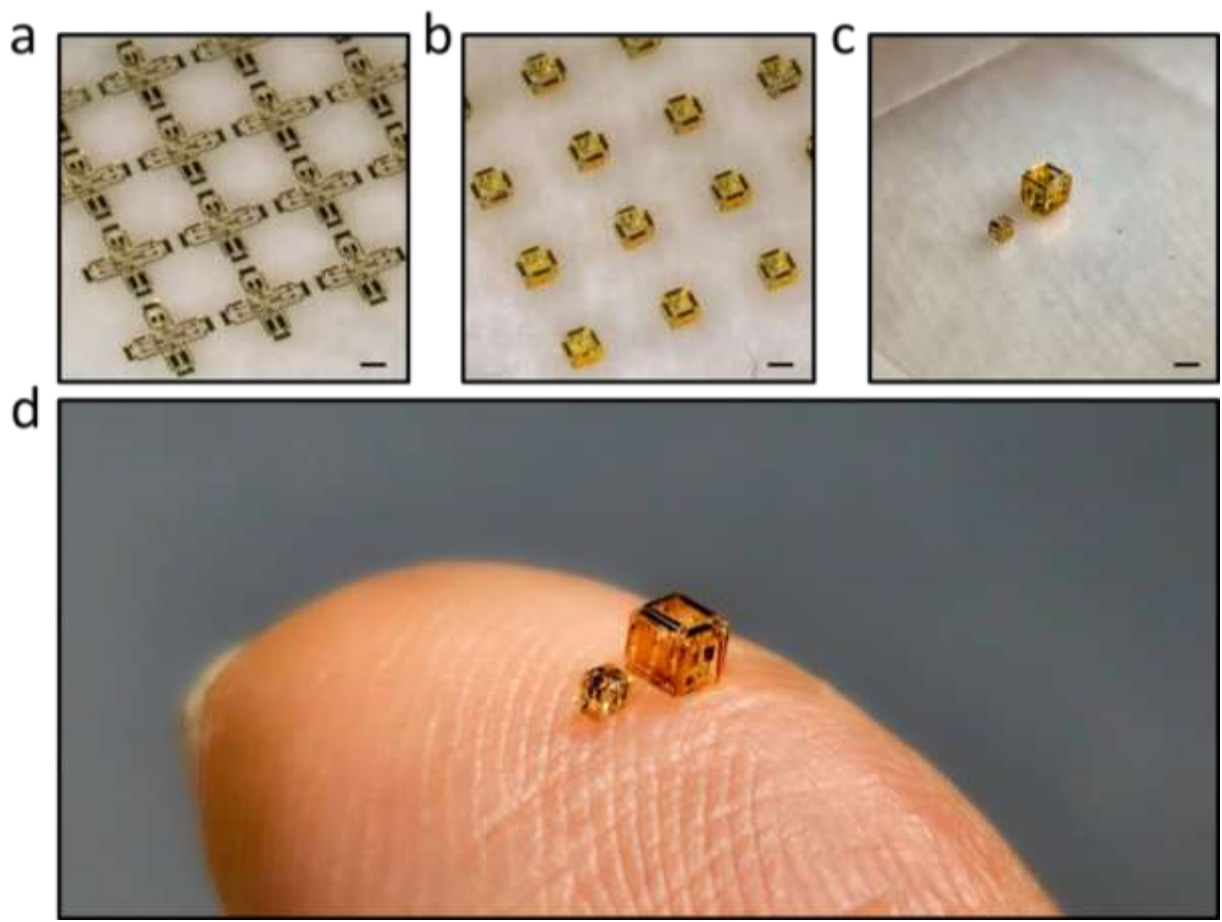

**Figure S2 | Two smartlets with volumes of 1 mm³ and 0.125 mm³. a,** Wafer-scale fabrication of smartlets with the edge length of 500 μm on a 2-inch substrate, showing highly parallel processing prior to self-assembly. **b,** Array of fully self-assembled smartlets with the edge length of 500 μm after release and folding. **c,** Down scaled smartlets with the edge length of 500 μm positioned on a 4-inch glass substrate next to generation-1 smartlets with the edge length of 1mm. **d,** Down scaled smartlets with the edge length of 500 μm and generation-1 smartlets with the edge length of 1 mm resting on a fingertip for scale comparison . Scale bars, 0.5 mm.